\documentstyle[preprint,aps]{revtex}

\tightenlines

\begin{document}


\title{Symmetry-preserving Loop Regularization and Renormalization of QFTs}
\author{Yue-Liang Wu }
 \address{Institute of Theoretical Physics, Chinese Academy of sciences,
 Beijing 100080, China  }
 %
\date{ylwu@itp.ac.cn}
\maketitle

\begin{abstract}
A new symmetry-preserving loop regularization method proposed in
\cite{ylw} is further investigated. It is found that its
prescription can be understood by introducing a regulating
distribution function to the proper-time formalism of irreducible
loop integrals. The method simulates in many interesting features
to the momentum cutoff, Pauli-Villars and dimensional
regularization. The loop regularization method is also simple and
general for the practical calculations to higher loop graphs and
can be applied to both underlying and effective quantum field
theories including gauge, chiral, supersymmetric and gravitational
ones as the new method does not modify either the lagrangian
formalism or the space-time dimension of original theory. The
appearance of characteristic energy scale $M_c$ and sliding energy
scale $\mu_s$ offers a systematic way for studying the
renormalization-group evolution of gauge theories in the spirit of
Wilson-Kadanoff and for exploring important effects of higher
dimensional interaction terms in the infrared regime.
\end{abstract}

\vspace{0.8cm} \hspace{1cm}{\bf PACS numbers}: 11.10.Cd,11.15.Bt

\hspace{0.8cm} {\bf Keywords:} Loop Regularization, Irreducible
loop integrals, Consistency conditions, Regulating distribution
function, Characteristic and Sliding energy scales

\newpage


 An important issue for making quantum field theories (QFTs)
to be physically meaningful is the elimination of ultraviolet (UV)
divergences without spoiling symmetries of the original theory.
This is because whether QFTs are underlying or effective theories,
they must be finite theories for describing the real world. To
avoid such difficulties, one may modify the behavior of field
theory at very large momentum or even at very small momentum if
there exist infrared (IR) divergences, so that all Feynman
diagrams become well-defined quantities. Such a procedure is
usually called regularization. There are numerous regularization
schemes which are available for making theory finite. The most
frequently adopted regularization schemes in perturbative
calculations include the momentum cutoff\cite{MC}, Pauli-Villars
procedure\cite{PV}, Schwinger proper-time scheme\cite{PT}, and
dimensional regularization\cite{DR}. However, not all of the
regularization schemes preserves all symmetries of the original
theory. In particular, the construction of a regularization that
respects to non-Abelian gauge symmetry has turned out to be a
difficult task. The naive momentum cutoff is known to destroy not
only the translational invariance but also the gauge invariance.
Alternatively, a higher covariant derivative Pauli-Villars
regularization was proposed \cite{AAS} to regulate all the
divergences and meanwhile respect to the gauge symmetry in
non-Abelian gauge theories. Nevertheless, such a regularization
must be carefully checked as it may lead to an inconsistent
quantum chromodynamics (QCD)\cite{MR}. The most popular gauge
symmetry-preserving regularization is the dimensional
regularization. While the method fails for the case in which the
system under consideration is specific to the initial space-time
dimension (e.g. $\gamma_5$ in four dimensions, chiral and
supersymmetric theories ), and also for the case in which the
scaling behavior becomes important. The well-known example for the
later case is the derivation of gap equation in the gauged
Nambu-Jona-Lasinio model. It was shown\cite{TG} that the
dimensional regularization cannot lead to a correct gap equation.
This is because the dimensional regularization destroys the
quadratic `cutoff' momentum term in the gap equation. For the same
reason, it fails to calculate the chiral loop contributions which
play an important role for understanding the $\Delta I = 1/2$ rule
and predicting direct CP violation $\varepsilon'/\varepsilon$ in
the $K \rightarrow \pi\pi$ decays\cite{ylwu}. In this sense, no
single satisfactory regularization can be applied to all purposes
in QFTs. Nevertheless, this does not influence the enormous
successful applications of QFTs in describing the real world. In
fact, according to the Weinberg's folk theorem which states
that\cite{SW1,SW2}: any quantum theory that at sufficiently low
energy and large distances looks Lorentz invariant and satisfies
the cluster decomposition principle will also at sufficiently low
energy look like a quantum field theory. This implies that there
must exist in any case a characteristic energy scale (CES) $M_c$
which can be either a fundamental-like energy scale (such as the
Planck scale $M_P$ or the string scale $M_s$ in string theory) or
a dynamically generated energy scale (for instance, the chiral
symmetry breaking scale $\Lambda_{\chi}$ and the critical
temperature for superconductivity), so that one can always make an
QFT description at a sufficiently low energy scale in comparison
with the CES $M_c$. On the other hand, basing on the spirit of
renormalization group by Wilson-Kadanoff\cite{WK} and
Gell-Mann-Low\cite{GML}, one should be able to deal with physical
phenomena at any interesting renormalization energy scale or the
so-called sliding energy scale (SES) $\mu_s$ by integrating out
the physics at higher energy scales. Therefore, the explicit
regularization method is expected to be governed by a physically
meaningful CES $M_c$ and a physically interesting SES $\mu_s$.
From the above analyzes, there should be no doubt of existing a
new symmetry-preserving and infinity-free regularization scheme
which can be realized in the space-time dimension of original
theory without modifying the original lagrangian formalism. This
belief has recently been explored in detail by introducing the
concept of irreducible loop integrals (ILIs)\cite{ylw}. In
general, n-fold ILIs that are evaluated from n-loop overlapping
Feynman integrals of loop momenta $k_i$ ($i=1,\cdots, n$) are
defined as the loop integrals in which there are no longer the
overlapping factors $(k_i-k_j + p_{ij})^2$ $(i\ne j)$ that appear
in the original overlapping Feynman integrals.

 It has been shown in \cite{ylw} that
a set of regularization independent consistency conditions  must
be satisfied for regularized ILIs to preserve gauge invariance of
the original theory. A general symmetry-preserving new
regularization prescription has been proposed to realize the
consistency conditions. To present an alternative and independent
verification, we shall demonstrate in this note that by applying
the Schwinger proper-time formalism to 1-fold ILIs the new
regularization prescription can well be understood through
constructing a regulating distribution function for the
proper-time integration of the 1-fold ILIs. As a consequence, the
gauge invariant nature becomes manifest in such a new
demonstration. This is because in the proper-time formalism the
divergence in momentum integration transfers into the singularity
of the proper-time variable $\tau$ which is independent of gauge
transformation. Unlike the Pauli-Villars scheme, our
regularization prescription is applied to the ILIs of Feynman loop
graphs rather than to the propagators as imposed in the
Pauli-Villars procedure. Also unlike the Pauli-Villars and
dimensional regularization, the new regularization prescription do
not modify either the lagrangian formalism or the space-time
dimension of original theory\footnote{It is noticed that a
constrained differential renormalization\cite{CDR} was also
proposed with one of the motivations that the original lagrangian
formalism and space-time dimension should not be modified.}. To be
distinguishable and convenient for further mention in the
following discussions, we may call such a consistent and
symmetry-preserving new regularization as a Loop Regularization
(LR). The paper is organized as follows, we shall start from the
consistency conditions and the general regularization prescription
for 1-fold ILIs of one-loop graphs, we then present a detailed
description for the construction of a regulating distribution
function and show how it can reproduce the general regularization
prescription proposed in \cite{ylw}, we finally illustrate for
completeness of demonstration how the consistency conditions and
regularization prescription can straightforwardly be generalized
to higher fold LILs of arbitrary loop graphs, and how a set of
theorems can practically be used to construct general proofs of
renormalizability in QFTs. In fact, the new symmetry-preserving LR
method has been found to be practically very useful and reliable
for deriving the chiral effective field theory and for describing
the dynamically spontaneous symmetry breaking for low energy
dynamics of QCD\cite{DWu}, where the light nonet scalar mesons
have been shown to play the role of composite Higgs bosons and the
resulting mass spectra for both scalar and pseudoscalar nonet
mesons are consistent with the current experimental data.

Let us begin with the one loop case. By adopting the Feynman
parameter method, all the one loop integrals can be expressed in
terms of the following 1-fold ILIs\cite{ylw}
    \begin{eqnarray}
    && I_{-2\alpha} = \int \frac{d^4 k}{(2\pi)^4}\ \frac{1}{(k^2 - {\cal M}^2)^{2+\alpha}}\
    ,  \nonumber \\
    && I_{-2\alpha\ \mu\nu} = \int \frac{d^4 k}{(2\pi)^4}\
    \frac{k_{\mu}k_{\nu}}{(k^2 - {\cal M}^2)^{3 + \alpha} }\ , \nonumber \\
    && I_{-2\alpha\ \mu\nu\rho\sigma} = \int \frac{d^4 k}{(2\pi)^4}\
    \frac{k_{\mu}k_{\nu} k_{\rho}k_{\sigma} }{(k^2 - {\cal M}^2)^{4+ \alpha} }
    \end{eqnarray}
with $\alpha =-1, 0, 1, \cdots$. Here $\alpha = -1$ and $\alpha =
0$ correspond to the quadratically ($I_2$, $I_{2 \mu\nu \cdots}$)
and logarithmically ($I_0$, $I_{0 \mu\nu \cdots}$) divergent
integrals. The mass factor ${\cal M}^2$ is in general a function
of the Feynman parameters and external momenta $p_j$. All the
Feynman parameter integrations are omitted in the paper.

The consistency conditions of gauge invariance for regularized
1-fold ILIs turn out to be in the four dimensional
space-time\cite{ylw}
\begin{eqnarray}
& & I_{2\mu\nu}^R = \frac{1}{2} g_{\mu\nu}\ I_2^R, \quad
I_{2\mu\nu\rho\sigma }^R = \frac{1}{8} g_{ \{\mu\nu}
g_{\rho\sigma\} }\ I_2^R  , \nonumber \\
& & I_{0\mu\nu}^R = \frac{1}{4} g_{\mu\nu} \ I_0^R, \quad
I_{0\mu\nu\rho\sigma }^R = \frac{1}{24} g_{ \{\mu\nu}
g_{\rho\sigma\} }\ I_0^R
\end{eqnarray}
with $g_{ \{\mu\nu} g_{\rho\sigma\} } \equiv
g_{\mu\nu}g_{\rho\sigma} + g_{\mu\rho}g_{\nu\sigma} +
g_{\mu\sigma}g_{\rho\nu}$. Where the superscript `R' denotes the
regularized ILIs. Once the above consistency conditions hold, the
divergent structure of theories can be well characterized by two
regularized scalar-type ILIs $I_0^R$ and $I_2^R$. For underlying
gauge theories, the regularized quadratic divergences are found to
cancel each other under the consistency conditions. One only needs
to consider the regularized logarithmic divergent ILIs $I_0^R$
which can fully be absorbed into the redefinitions of coupling
constants and relevant fields. Thus the consistency conditions
enable one to establish a well-defined renormalizable and gauge
invariant theories without introducing any specific regularization
scheme. In this sense, we arrive at a regulator-free scheme for
underlying quantum gauge field theories. Therefore the consistency
conditions become the key criteria for constructing an explicit
gauge symmetry-preserving regularization scheme. As a fact, one
can easily show from a general Lorentz structure analysis that
$I_{2\mu\nu}= a g_{\mu\nu}I_2$ with $a =
I_{2\mu\nu}g^{\mu\nu}/(4I_2) =1/4$, which does not match to the
consistency conditions and will spoil gauge invariance. In
contrast, when considering only the time component, i.e., $I_{2\
00}= a g_{00} I_2$, as the integration over the component $k_0$
are convergent, one can perform the integration over $k_0$ for
both $I_{2\ 00}$ and $I_2$, which leads to $a = 1/2$ that
coincides with the consistency conditions. We are then led to a
general {\bf theorem} that {\it the convergent integrations can
safely be carried out, only the divergent integrations destroy the
gauge invariance.}

It is of interest to note that a more general regularization
scheme that ensures the consistency conditions can truly be
realized via a simple prescription\cite{ylw}, that is: in the four
dimensional Euclidean space of momentum, replacing in the ILIs the
loop integrating variable $k^2$ and the loop integrating measure
$\int d^4 k$ by the corresponding regularized ones $[k^2]_l$ and
$\int [d^4 k]_l$
\begin{eqnarray}
& &  \quad k^2  \rightarrow [k^2]_l \equiv k^2 + M^2_l\ ,
\nonumber \\
& & \int d^4 k \rightarrow \int [d^4 k]_l \equiv \lim_{N, M_l^2}
\sum_{l=0}^{N} c_l^N \int d^4k
\end{eqnarray}
where $M_l^2$ ($ l= 0,1,\ \cdots $) are the mass factors of
regulators. One takes $M_0^2 = 0$  and $ c_0^N = 1$ to maintain
the original integrals. For IR divergent integrals, one can set
$M_0^2 = \mu_s^2$ to avoid it. The coefficients $c_l^N$ are chosen
to modify the short-distance behavior of loop integrals and to
remove the divergences via the following conditions
\begin{eqnarray}
\lim_{N, M_l^2}\sum_{l=0}^{N} c_l^N\ (M_l^2)^n = 0 \quad
       (n= 0, 1, \cdots)
\end{eqnarray}
where $\lim_{N, M_l^2} \equiv \lim_{N, M_l^2\rightarrow \infty}$.
Thus all regularized high order integrals vanish, i.e., $\int [d^4
k]_l \ (k^2 + M_l^2)^n =0 $.

As an explicit and simple solution of eq.(4), it is not difficult
to find the following interesting solution
\begin{equation}
M_l^2 = \mu_s^2 + l M_R^2, \quad c_l^N = (-1)^l \frac{N!}{(N-l)!\
l!}
\end{equation}
Here $M_R$ is an arbitrary mass scale. With the above solution,
the regularized ILIs $I_0^R$ and $I_2^R$ get explicit
forms\cite{ylw}
\begin{eqnarray}
I_2^R  & = & \frac{-i}{16\pi^2} \ \{\ M_c^2 - \mu^2 [ \ \ln
\frac{M_c^2}{\mu^2} - \gamma_w + 1 +   y_2(\frac{\mu^2}{M_c^2})  \ ] \ \}  \nonumber \\
I_0^R  & = & \frac{i}{16\pi^2} \ [ \ \ln \frac{M_c^2}{\mu^2 } -
\gamma_w + y_0(\frac{\mu^2}{M_c^2}) \ ]
\end{eqnarray}
with $ \mu^2 = \mu_s^2 + {\cal M}^2$, $\gamma_w = \gamma_E =
0.5772\cdots$, and
\begin{eqnarray}
& & y_0 (x) = \int_0^x d \sigma\ \frac{1 - e^{-\sigma} }{\sigma},
\quad  y_1 (x)  = \frac{e^{-x} - 1 + x}{x} \nonumber \\
& & y_2(x) = y_0(x) - y_1(x),\quad  M_c^2 = \lim_{N,M_R} M_R^2/\ln
N
\end{eqnarray}
One can compare the above results with the ones in the momentum
cutoff and dimensional regularization. It is easily seen that
$\mu_s $ sets an IR `cutoff' at ${\cal M}^2 =0$ and $M_c$ provides
an UV `cutoff'. More generally speaking, $M_c$ and $\mu_s$ play
the role of CES and SES respectively. As such a new
symmetry-preserving LR method maintains the quadratic `cutoff'
term $M_c^2$, which makes it analogous to the momentum cutoff and
can be applied to effective QFTs. Meanwhile, it preserves the
symmetry principles as the dimensional regularization does. Of
particular, the divergent behaviors of the original theory are
recovered by simply taking $M_c\rightarrow \infty$. For underlying
renormalizable gauge theories, all the divergences at
$M_c\rightarrow \infty$ can be absorbed into the renormalization
of coupling constants and quantum fields. Before proceeding, we
want to point out that the prescription eq.(3) formally appears to
be similar to the one in Pauli-Villars procedure, but they are
conceptually quite different. This is because the prescription
here is applied to the ILIs of Feynman loop graphs rather than to
the propagators in the original Pauli-Villars procedure which is
known to spoil the gauge invariance (especially in non-Abelian
gauge theory). In addition, it is unlike the Pauli-Villars scheme
in which the original Lagrangian formalism is considered to be
modified with the introduction of nonphysical fields, and also
unlike the dimensional regularization in which the space-time
dimension of original theory has been changed by an analytic
continuation, while in the new symmetry-preserving LR method, both
the Lagrangian formalism and space-time dimension of original
theory are unchanged, only the divergent behavior has been
modified so that it becomes well-defined in the new LR method.
Furthermore, as shown in the new LR method, only when taking the
regulator number $N$ to be infinitely large, the resulting
regularized theory can become regulator-independent.

We now turn to the main issue in this paper that how the above
regularization prescription can be understood by constructing a
regulating distribution function. For demonstration, applying the
Schwinger proper-time formalism to the scalar type ILIs which are
rotated into the Euclidean momentum space
\begin{eqnarray}
& & I_{-2\alpha} = i(-1)^{\alpha} \int \frac{d^4 k}{(2\pi)^4}\
\frac{1}{(k^2 + {\cal M}^2)^{2+\alpha}}  \nonumber \\
& & = \frac{i(-1)^{\alpha}}{\Gamma(\alpha +2 )} \int \frac{d^4
k}{(2\pi)^4}\ \int_0^{\infty} d\tau\ \tau^{\alpha + 1} e^{-\tau
(k^2 + {\cal M}^2)}
\end{eqnarray}
where the identity $a^{-n} \Gamma(n) = \int_0^{\infty} dx\ x^{n-1}
e^{-ax}$ has been used. For illustration, consider the divergent
ILIs $I_0$. By exchanging the order of integrations for the
momentum $k$ and the `proper-time' variable $\tau$, and carrying
out safely the integration over $k$ as it becomes convergent
\begin{eqnarray}
I_{0} = \frac{i}{16\pi^2}   \int_0^{\infty} \frac{d\tau}{\tau}\
e^{-\tau {\cal M}^2}
\end{eqnarray}
one sees that the UV divergence for the momentum integration
transfers into the singularity for the proper-time integration at
$\tau=0$. At ${\cal M}^2 =0$, the IR divergence for the momentum
integration also transfers into the singularity for the
proper-time integration at $\tau= \infty$. Therefore in the
proper-time formalism, any possible divergence originating from
momentum integration will be turned into the singularity in
$\tau$. Thus instead of making a regularization on the momentum
integration, one can choose the regularization procedure operating
on the proper-time integration of ILIs by multiplying a
$R^{\infty}$ regulating distribution function ${\cal  W}_N(\tau;
M_c, \mu_s)$. Here $M_c$ and $\mu_s$ are two scales which play the
role of UV and IR cutoff scales respectively. In general, the
regulating distribution function ${\cal  W}_N(\tau; M_c, \mu_s)$
must satisfy four conditions:

(i) ${\cal W}_N(\tau=0; M_c, \mu_s) = 0$ so as to eliminate the
singularity at $\tau = 0$ which corresponds to the UV divergence
for the momentum integration.

(ii) ${\cal  W}_N(\tau=\infty; M_c, \mu_s=0) = 1$ for ensuring the
regulating distribution function not to modify the behavior of
original theory in the IR regime.

(iii) ${\cal W}_N(\tau; M_c\rightarrow \infty, \mu_s=0) = 1$ which
ensures the proper-time formalism to recover the original ILIs in
the physical limits.

(iv) $\lim_{N\rightarrow \infty} {\cal W}_{N} (\tau; M_c, \mu_s=0)
= 1$  for $\tau \geq 1/M_c^2$ and $\lim_{N\rightarrow \infty}
{\cal  W}_{N} (\tau; M_c, \mu_s=0) = 0$  for $\tau < 1/M_c^2$, so
that $M_c$ acts as the UV cutoff scale.

The regularized proper-time formalism for ILIs is defined by
\begin{eqnarray}
I_{-2\alpha}^R & & = \frac{i(-1)^{\alpha}}{\Gamma(\alpha +2 )}
 \lim_{N\rightarrow \infty} \int \frac{d^4 k}{(2\pi)^4}\ \nonumber \\
& & \int_0^{\infty} d\tau\ {\cal  W}_{N}(\tau; M_c, \mu_s)\
\tau^{\alpha + 1} e^{-\tau (k^2 + {\cal M}^2)}
\end{eqnarray}
The simple choice for the function ${\cal  W}_{N}(\tau; M_c,
\mu_s)$ is
\begin{eqnarray}
{\cal  W}_{N}(\tau; M_c, \mu_s) =  e^{-\tau \mu_s^2} \left(1 -
e^{-\tau M_R^2} \right)^N
\end{eqnarray}
which satisfies the required four conditions for $M_R^2 = M_c^2
h_w(N)\ln N$ with $h_w(N)\agt 1$ and $h_w(N\rightarrow \infty)
=1$. Rewriting $ {\cal W}_{N}(\tau; M_c, \mu_s)$ into the
following form
\begin{eqnarray}
{\cal  W}_{N}(\tau; M_c, \mu_s) = \sum_{l=0}^{N} (-1)^l
\frac{N!}{(N-l)!\ l!}\  e^{-\tau(\mu_s^2 + lM_R^2)}
\end{eqnarray}
and substituting it into eq.(10), we then obtain, by integrating
over $\tau$, the following interesting result
\begin{eqnarray}
I_{-2\alpha}^R & = & \lim_{N, M_l^2} \int \frac{d^4 k}{(2\pi)^4}\
\sum_{l=0}^{N} c_l^N \
\frac{i(-1)^{\alpha}}{(k^2  + {\cal M}^2 + M_l^2)^{\alpha + 2}}\nonumber \\
& = & i(-1)^{\alpha} \int \frac{[d^4 k]_l}{(2\pi)^4}\
\frac{1}{([k^2]_l  + {\cal M}^2 )^{\alpha + 2}}
\end{eqnarray}
which exactly reproduces the prescription described in eqs.(3-5)
for the new symmetry-preserving LR method. In here the gauge
invariant nature is manifest as in the proper-time formalism the
divergence in momentum integration transfers into the singularity
of the proper-time variable $\tau$ which is independent of gauge
transformation.

We now briefly discuss the case with more closed loops. Though the
proper-time scheme of 1-fold ILIs with a simple regulating
distribution function can exactly reproduce the prescription
described in eqs.(3-5) for 1-fold ILIs, while it will be seen that
when applying the regularization method to more closed loops, the
generalization of the prescription described in eqs.(3-5) and
reproduced by the regulating distribution function in the
proper-time formalism of 1-fold ILIs (see eq.(13)) becomes much
more manifest and straightforward than the generalization of the
proper-time scheme in which one needs first adopt a proper-time
formalism to higher fold ILIs and then construct a regulating
distribution function for the proper-time integration of the
higher fold ILIs. As it has been shown in \cite{ylw} that any loop
integrals can be evaluated into the corresponding ILIs by
repeatedly using the Feynman parameter method and the
UV-divergence preserving parameter method
\begin{equation}
\frac{1}{a^{\alpha}b^{\beta}} = \frac{\Gamma(\alpha +
\beta)}{\Gamma(\alpha)\Gamma(\beta)} \int_{0}^{\infty}\ du
\frac{u^{\beta -1} }{[a + b u]^{\alpha +\beta } }
\end{equation}
Here $u$ is the UV-divergence preserving integral variable. As a
consequence, the UV divergence for the momentum integration
transfers into the one for $u$ integration. Explicitly, n-fold
ILIs are found to have the following general form after safely
performing (n-1) convergent integrations over the momentum $k_i$
($i=1,\cdots, n-1$),
\begin{eqnarray}
& & I^{(n)}_{\Delta} =  \prod_{i=1}^{n-1} \int_0^{\infty} du_i
\frac{F_{is}(x_{lm})}{(u_i +
\rho_i)^{\Delta_{is}} }\  I^{(1)}_{\Delta_n}(\mu_n^2) \\
& & I^{(n)}_{\Delta \mu\nu } = \prod_{i=1}^{n-1} \int_0^{\infty}
du_i \frac{F_{is}(x_{lm})}{(u_i + \rho_i)^{\Delta_{is}} } \
I^{(1)}_{\Delta_n \mu\nu }(\mu_n^2) \nonumber
\end{eqnarray}
with
\begin{eqnarray}
& & I^{(1)}_{\Delta_n}(\mu_n^2) = \int d^4 k_n \frac{1}{(k_n^2
+ {\cal M}^2 + \mu_n^2)^{\Delta_n}  } \\
& & I^{(1)}_{\Delta_n \mu\nu }(\mu_n^2)  = \int d^4 k_n
\frac{k_{n\mu}k_{n\nu}}{(k_n^2 + {\cal M}^2 + \mu_n^2)^{\Delta_n +
1} } \nonumber
\end{eqnarray}
Here we have only presented the lowest order tensor-type ILIs, its
generalization to higher order tensor-type ILIs is
straightforward. Where $u_i$ ($i=1,\cdots n-1$) are the
UV-divergence preserving integral variables and $x_{lm}$ ($l,m =1,
2, \cdots$) the usual Feynman parameters. For a given overlapping
Feynman loop integral, $F_{is}(x_{lm})$ are the known functions of
Feynman parameters, $\Delta_{is}$ and $\Delta_n$ are the known
powers, $\rho_i$ and $\mu_n^2$ are also the known functions of
$u_i$ and $x_{lm}$ with the property: $\rho_1 = \rho_1(x_{1m})$,
$\rho_i = \rho_i(x_{lm},\ u_1, \cdots u_{i-1})$ ($i =2, \cdots
n-1$, $l=1, \cdots i$) and $\mu_n^2 = \mu_n^2(x_{lm},\
u_1,\cdots,u_{n-1})$. It is remarkable to observe that the
functions $\rho_i$ and $\mu_n^2$ have a vanishing limit at the
divergent points, namely
\begin{eqnarray}
\rho_i \rightarrow 0, \quad \mu_n^2 \rightarrow 0 \quad \mbox{for}
\quad u_{1},\cdots,u_{n-1} \rightarrow \infty
\end{eqnarray}
From this important feature and the general form of the n-fold
ILIs, we can verify\cite{ylw} that in the resulting n-fold ILIs
the integral over the n-th loop momentum $k_n$ describes the
overall divergent property of n-loop diagrams and the
sub-integrals over the variables $u_i$ ($i=1,2,\cdots, n-1$)
characterize the UV divergent properties for the one-loop,
two-loop, $\cdots$, (n-1)-loop sub-diagrams respectively. With
these observations, we arrive at a key {\bf theorem} that {\it in
the Feynman loop integrals the overlapping divergences which
contain overall divergences and also divergences of sub-integrals
will become factorizable in the corresponding ILIs}. Based on all
the features, we are led to a main {\bf theorem} that {\it all the
overlapping divergent integrals can be made to be harmless via
appropriate subtractions}. From the tensor-type n-fold ILIs, one
notices that the tensor structure is actually characterized by the
overall 1-fold ILIs $I^{(1)}_{\Delta_n\ \mu\nu }(\mu_n^2)$, thus
the consistency conditions for one loops can straightforwardly be
generalized to any fold ILIs of arbitrary loops. In fact, we can
deduce a more general statement or {\bf theorem} that {\it the
consistency conditions between the tensor and scalar type overall
1-fold ILIs are necessary and sufficient for ensuring the gauge
invariance of gauge theories }.

 The generalization of the regularization
prescription in eqs.(3-5) to any higher fold ILIs becomes
straightforward, that is \cite{ylw}: universally replace in the
n-fold ILIs the n-th loop momentum square $k^2_n$ and the
corresponding loop integrating measure $\int d^4 k_n$ as well as
the UV-divergence preserving integral variables $u_i$ ($i=1,
\cdots, n-1$) and the corresponding integrating measure $\int d
u_i $ by the regularizing ones $[k^2_n]_l$ and $\int [d^4 k_n]_l$
as well as $[u_i]_l$  and $\int [du_i]_l$ with $n$ being
arbitrary. Explicitly, one has
\begin{eqnarray}
k^2_n & \rightarrow & [k^2_n]_l \equiv k^2_n + M^2_l =  k^2_n + \mu_s^2 + l M^2_R \ ,\nonumber \\
\int d^4 k_n &  \rightarrow  & \int [d^4 k_n]_l \equiv \lim_{N,
M_R^2} \sum_{l=0}^{N} c_l^N \int d^4 k_n   \nonumber \\
u_i & \rightarrow & [u_i]_l \equiv u_i + M_l^2/\mu_s^2 =
u_i + 1 + l M_R^2/\mu_s^2   \nonumber \\
\int du_i & \rightarrow & \int [du_i]_l \equiv \lim_{N,M_R}
\sum_{l=0}^{N} c_l^N \int du_i
\end{eqnarray}
which, together with the solution of $c_l^N$ in eq.(5), present
the whole regularization prescription for the new
symmetry-preserving LR method.

It becomes clear that to regularize higher fold ILIs of more
closed loops, the generalization of the regularization
prescription presented in eqs.(3-5) and reproduced by the
regulating distribution function in eq.(13) is much more
straightforward than the generalization of proper-time scheme.
This is because the first step of expressing higher fold ILIs into
proper-time formalism is no longer as manifest as the one for the
case of 1-fold ILIs due to the overlapping integrals.

 For completeness, we illustrate in a more concise way
how the new symmetry-preserving LR method is practically a useful
means to handle overlapping divergences. A more detailed
verification has been described in \cite{ylw}. As a simple
example, we represent here an explicit treatment to two-loop
overlapping divergences. For comparison, the demonstration is made
along the line in analogous to the dimensional
regularization\cite{DR}. As shown in \cite{DR}, the scalar type
loop integrals in the general two loop diagrams can always been
expressed into the following overlapping integrals by using the
Feynman parameter method
     \begin{eqnarray}
     I_{\alpha\beta\gamma}^{(2)} = \int d^4 k_1 \int d^4 k_2
     \frac{1}{(k_1^2 +  {\cal M}_1^2 )^{\alpha}\left(k_2^2 + {\cal M}_2^2\right)^{\beta}
     \left((k_1-k_2 +p )^2 + {\cal M}_{12}^2\right)^{\gamma} }
     \end{eqnarray}
     which is the so-called general $\alpha\beta\gamma$ diagrams
     with $\alpha,\ \beta,\ \gamma > 0 $. The mass factors ${\cal M}_1^2$, ${\cal M}_2^2$ and ${\cal M}_{12}^2$
     are in general the functions of masses $m_i^2$ and external momenta $p_i^2$ ($i=1,2,
     \cdots$). Following the definitions in the ref.\cite{DR}:
     (i) the sub-integral over $k_1$ is said to be
     convergent or divergent according to $\alpha + \gamma > 2$ or $\alpha + \gamma \le 2$,
     similarly, the sub-integral over $k_2$ is said to be
     convergent or divergent according to $\beta + \gamma > 2$ or $\beta + \gamma \le 2$;
     (ii) the overall integral of  $\alpha\beta\gamma $ diagram is
     said to be overall convergent or overall divergent according to
     $\alpha + \beta + \gamma > 4$ or $\alpha + \beta + \gamma \le 4$;
     (iii) a harmless divergence
     is a divergence with its coefficient functions a polynomial of finite order in the
     external momenta. In the gauge theories, the overall divergence of a nontrivial
     overlapping integral is at most quadratic, thus $\alpha + \beta + \gamma \ge 3$.

     Repeatedly adopting the Feynman parameter method and the
     UV-divergence preserving method, we can arrive at the corresponding ILIs
     for the general $\alpha\beta\gamma$ diagrams
     \begin{eqnarray}
     & & I_{\alpha\beta\gamma}^{(2)}  =
     \Gamma_{\alpha \beta \gamma }\int_0^1 dx\ x^{\gamma -1}(1-x)^{\alpha -1}
     \int_0^{\infty} d u \frac{\pi^2 \ u^{\beta - 1}}{
     (u + x(1-x))^{\alpha + \beta + \gamma - 2} }\  I_{\alpha\beta\gamma}^{(1)}( \mu_{2}^2)  \nonumber  \\
     & & I_{\alpha\beta\gamma}^{(1)}( \mu_{2}^2) =  \int d^4 k_2 \frac{1}{(\ k_2^2 + {\cal M}_2^2
     + \mu_{2}^2 \ )^{\alpha + \beta + \gamma -2 } }
     \end{eqnarray}
     where we have introduced the definitions
     \begin{eqnarray}
     & & \Gamma_{\alpha \beta \gamma } =
     \frac{\Gamma(\alpha + \beta
     + \gamma - 2)}{\Gamma(\alpha)\Gamma(\beta)\Gamma(\gamma)}  \\
    & &  \mu_{2}^2 \equiv \mu_{2}^2(x,u)=
     \frac{1}{u + x(1-x) } \ \left(  {\cal M}^2(x,p^2) - \frac{x^2 (1-x)^2 }{u + x(1-x)}\  p^2 \
    \right) \\
    & & {\cal M}^2(x,p^2) = (1-x){\cal M}_1^2 + x {\cal M}_{12}^2
     -x(1-x) (\ {\cal M}_2^2 - p^2\ )
     \end{eqnarray}
As the external momentum dependence in the two-fold ILIs only
appears in the mass factor ${\cal M}_2^2  + \mu^2_2$ of the
overall one-fold ILI $I_{\alpha\beta\gamma}^{(1)}( \mu_{2}^2)$,
and the sub-integral over the variable $u$ preserves one-loop UV
divergent structure over the loop momentum $k_1$, one only needs
to introduce one subtraction term of the sub-integral over the
loop momentum $k_1$
  \begin{eqnarray}
     I_{\alpha\beta\gamma}^{(2)S} = \int d^4 k_1
     \frac{1}{(k_1^2 + m_o^2 )^{\alpha + \gamma} }
     \int d^4 k_2 \frac{1}{\left(k_2^2 + {\cal M}_2^2\right)^{\beta} } \ ,
  \end{eqnarray}
where the superscript $S$ denotes the subtraction term. As the
feature: $\mu_2^2  \rightarrow 0$ at $u \rightarrow \infty $, the
difference of the integrals, i.e., ( $I_{\alpha\beta\gamma}^{(2)}
- I_{\alpha\beta\gamma}^{(2)S}$ ), contains only harmless
divergences.

   To be more clear, applying the general prescription of new symmetry-preserving LR regularization to
   the ILIs, we then obtain the well-defined regularized two-fold ILIs
   \begin{eqnarray}
     & & I_{\alpha\beta\gamma}^{(2)R}  =
     \Gamma_{\alpha \beta \gamma }\int_0^1 dx\ x^{\gamma -1}(1-x)^{\alpha -1}
     \int_0^{\infty} [d u]_l \frac{\pi^2 \ [u]_l^{\beta - 1}}{
     ([u]_l + x(1-x))^{\alpha + \beta + \gamma - 2} }\  I_{\alpha\beta\gamma}^{(1)R}( [\mu_{2}^2]_{l'})  \nonumber  \\
     & & I_{\alpha\beta\gamma}^{(1)R}( [\mu_{2}^2]_l) =  \int [d^4 k_2]_{l} \frac{1}{(\  [k_2^2]_{l} + {\cal M}_2^2
     + [\mu_{2}^2]_l \ )^{\alpha + \beta + \gamma -2 } }
     \end{eqnarray}
and the corresponding regularized subtraction term
  \begin{eqnarray}
     I_{\alpha\beta\gamma}^{(2)RS} & = & \int [ d^4 k_1 ]_l
     \frac{1}{( [k_1^2]_l + m_o^2 )^{\alpha + \gamma} }
     \int [ d^4 k_2 ]_{l'} \frac{1}{\left( [k_2^2]_{l'} + {\cal M}_2^2 \right)^{\beta} }
     = I_{\alpha\gamma}^{(1)R} I_{\beta}^{(1)R}
  \end{eqnarray}

According to the factorization and subtraction theorems for the
overlapping divergences as well as the harmless divergence
theorem, we shall be able to express the regularized ILIs into the
following general form
     \begin{eqnarray}
     I_{\alpha\beta\gamma}^{(2)R} = I_{\alpha\gamma}^{(1)RD}I_{\alpha\beta\gamma}^{(1)RD}
     + I_{\alpha\gamma}^{(1)RC}I_{\alpha\beta\gamma}^{(1)RD}
     + I_{\alpha\beta\gamma}^{(2)RC}
     \end{eqnarray}
where the superscripts `$RD$' and `$RC$' represent the regularized
divergent and convergent ILIs respectively, and the numbers (1)
and (2) in the superscripts label the one-fold and two-fold ILIs
respectively. We shall show that the first term with double
divergences is harmless after an appropriate subtraction. As the
regularized ILIs are well behaved, we are able to check the
validity of the above decomposition by explicitly carrying out the
integrations. To be more explicit, we consider two cases: (i)
$\alpha + \beta + \gamma = 4$ with $\alpha + \gamma = 2$ and
$\beta = 2$. From the usual power counting rule, the overall
integral in this case is logarithmically divergent ($\alpha +\beta
+\gamma =4$) and the sub-integral over $k_1$ is also
logarithmically divergent ($\alpha +\gamma =2$); (ii) $\alpha +
\beta + \gamma = 3$ with $\alpha + \gamma = 2$ and $\beta = 1$. In
this case, the overall integral is quadratically divergent, while
sub-integrals over $k_1$ ($\alpha\gamma$) and $k_2$
($\beta\gamma$) are logarithmically divergent.

 After performing some integrations over $u$ and $k$, we
 find for the case (i) and (ii) that
 \begin{eqnarray}
    & &  I_{0}^{(2)R} = \hat{I}_{0}^{(1)RD} I_{0}^{(1)RD} + \hat{I}_{0}^{(1)RC}I_{0}^{(1)RD}
     + I_{0}^{(2)RC}  \\
    & &  I_{2}^{(2)R} = \hat{I}_{0}^{(1)RD} I_{2}^{(1)RD} + \hat{I}_{2}^{(1)RC} I_{2}^{(1)RD} +
     I_{2}^{(2)RC}
     \end{eqnarray}
    with
   \begin{eqnarray}
    \hat{I}_{0}^{(1)RD} & = & \pi^2\int_0^1 dx\ [\ \ln \frac{M_c^2}{\mu_{x}^2 }
     - \gamma_w + y_0(\frac{\mu_{x}^2}{M_c^2})  \ ]  \\
     I_{0}^{(1)RD} & = & \pi^2 [\ \ln \frac{M_c^2}{\mu_M^2 }
     - \gamma_w + y_0(\frac{\mu_M^2}{M_c^2}) \ ]  \\
     \hat{I}_{0}^{(1)RC} & = & -\pi^2\int_0^1 dx\  \frac{x(1-x)}{ 1 + x -x^2}
     [\ 1 - y_{-2} (\frac{\mu_{x}^2}{M_c^2}) \ ] \\
    I_{0}^{(2)RC} & = & - \pi^4 \int_0^1 dx\  \int  [ d u]_l \
     \frac{ [u]_l }{(\  [u]_l + x(1-x)\ )^2 } \nonumber \\
    & &  [ \ \ln(\ 1 +  [\mu^2_2]_l/\mu_M^2\ )
     - z_0( [\mu^2_2]_l ) \ ] \\
 I_{2}^{(1)RD} & = & \pi^2\ \{\ M_c^2 - \mu_M^2\ [ \ \ln \frac{M_c^2}{\mu_M^2 }
     - \gamma_w + 1 +  y_2(\frac{\mu_M^2}{M_c^2}) \ ] \} \\
   \hat{I}_{2}^{(1)RC} & = & -\pi^2\  \int_0^1 dx\  \int [ d u ]_l \
     \frac{ [\mu^2_2]_l}{(\ [u]_l + x(1-x) \ ) } \\
 I_{2}^{(2)RC} & = &  \pi^4 \int_0^1 dx\ \int [ d u]_l \
     \frac{\mu_M^2 + [\mu^2_2]_l}{(\ [u]_l
     + x(1-x) \ ) } \nonumber \\
     & & [ \ \ln(\ 1 + [\mu^2_2]_l/\mu_M^2\ )
     - z_2([\mu^2_2]_l) \ ]
  \end{eqnarray}
 where
  \begin{eqnarray}
  & &  \mu_{x}^2 = (1 + x - x^2)\mu_s^2 \ , \qquad \mu_M^2 = \mu_s^2 +  {\cal
   M}_2^2 \\
   & & z_0([\mu^2_2]_l ) =  y_0(\frac{\mu_M^2 + [\mu^2_2]_l  }{M_c^2})
    - y_0(\frac{\mu_M^2}{M_c^2}) \\
  & &   z_2([\mu^2_2]_l ) =  y_2(\frac{\mu_M^2 + [\mu^2_2]_l  }{M_c^2})
    - y_2(\frac{\mu_M^2}{M_c^2})
  \end{eqnarray}

 The regularized subtraction term for the two cases can simply be written as
 \begin{eqnarray}
    & &  I_{0}^{(2)RS} = \hat{I}_{0}^{(1)RD}(m_o) I_{0}^{(1)RD} \\
     & &  I_{2}^{(2)RS} = \hat{I}_{0}^{(1)RD}(m_o) I_{2}^{(1)RD}
  \end{eqnarray}
with $m_o$ being the chosen subtraction point. Thus the
differences
  \begin{eqnarray}
    & & I_{0}^{(2)R}- I_{0}^{(2)RS} =  \left(\tilde{I}_0^{(1)RC}
    + \hat{I}_{0}^{(1)RC}\right) I_{0}^{(1)RD} + I_{0}^{(2)RC}  \\
    & &  I_{2}^{(2)R} -I_{2}^{(2)RS} = \left(\tilde{I}_0^{(1)RC}
    + \hat{I}_{2}^{(1)RC} \right) I_{2}^{(1)RD} + I_{2}^{(2)RC}
    \end{eqnarray}
  contain only harmless divergencies at $M_c \rightarrow \infty $. Here $\tilde{I}_0^{(1)RC}$
  is the additional convergent function arising from the subtraction
  \begin{eqnarray}
  \tilde{I}_0^{(1)RC} = \pi^2\int_0^1 dx\ [\ \ln \frac{m_o^2}{\mu_{x}^2 } +
     y_0(\frac{\mu_{x}^2}{M_c^2}) - y_0(\frac{m_o^2}{M_c^2}) \ ]
   \end{eqnarray}
The above explicit calculations demonstrate a practical
application of the new symmetry-preserving LR method in treating
the overlapping divergencies. Such prescriptions can
straightforwardly be generalized to more closed loops.

In conclusion, we have consistently demonstrated how the new
symmetry-preserving LR method\cite{ylw} of QFTs can be understood
by constructing a regulating distribution function in the
proper-time formalism of ILIs. An explicit regulating distribution
function has been obtained to reproduce the general regularization
prescription proposed in \cite{ylw}, which provides an alternative
verification and an independent check for the new
symmetry-preserving LR method. The advantages of the new
symmetry-preserving LR method become manifest that its
generalization to higher fold ILIs of more closed loops is
straightforward and its application for the practical calculations
is simple and general. Of particular, the presence of two energy
scales $M_c$ and $\mu_s$ allows us to study the
renormalization-group evolution of gauge theories in the spirit of
Wilson-Kadanoff scheme and to explore important effects of higher
dimensional new interaction terms characterized by the inverse
powers of $\mu_s$. Furthermore, the new symmetry-preserving LR
method appears to provide deep insights for dynamically
spontaneous symmetry breaking of strong interactions and mass
generation, and lead to profound applications for QFTs including
gauge, chiral, supersymmetric and gravitational ones, and also for
QFTs beyond four dimensional space-time (or with extra
dimensions). This is because the new symmetry-preserving LR method
is simply realized without modifying the original Lagrangian
formalism and is directly performed in the space-time dimension of
original theory. A more detailed study and application of the
symmetry-preserving new LR method is further in progress.

\centerline{\bf Acknowledgement}

 This work was supported in part by the projects of NSFC and CAS,
China and also by the associate scheme at the Abdus Salam
I.C.T.P., Italy.

\end{document}